# THE TRANSITION FROM GENERATION-RECOMBINATION NOISE IN BULK SEMICONDUCTORS TO DISCRETE SWITCHING IN SMALL-AREA SEMICONDUCTORS

FERDINAND GRÜNEIS

*Institute for Applied Stochastic*
*Rudolf von Scholtz Str. 4, 94036 Passau, Germany*
*Email: Ferdinand.Grueneis@t-online.de*

Scaling down the dimensions of Metal-Oxide-Semiconductor Field-Effect Transistors (MOSFETs) evinces discrete switching usually referred to as random telegraph signal (RTS). Such RTSs are usually attributed to the capture and emission of charge carriers by a single active trap located in an oxide layer. Machlup calculated the noise spectrum caused by these charge carriers based on probabilistic arguments. In this paper, we derive Machlup's noise spectrum differently: the g-r noise is attributed to a random succession of elementary g-r pulses. This enables g-r bulk noise to be interpreted in terms of the numbers of traps. The transition from g-r bulk noise to discrete switching is found by reducing the number of traps to just one single active trap. The resulting g-r noise spectrum is shown to be equivalent to Machlup's noise spectrum. The probability of an overlap of succeeding g-r pulses is calculated. Such an overlap is attributed to occupation of an empty single trap by an electron transferred from a neighboring trap. We simulate a g-r pulse train and find a large variety of patterns similar to those observed in MOSFETs. Excluding overlapping g-r pulses, the up-and-down distribution of succeeding g-r pulses is estimated.

*Keywords:* Generation-recombination noise, Small-area semiconductors, MOSFETs, Noise processes and phenomena in electronic transport, Statistical physics.

## 1. Introduction

The scaling down of device dimensions is accompanied by discrete switching of the device resistance, usually referred to as random telegraph signals (RTS). This phenomenon has been observed in different physical systems, such as Metal-Oxide-Semiconductor Field-Effect Transistors (MOSFETs), semiconductor nanowires, quantum dots, and others [1-11].

Reducing the channel dimension of a MOSFET, one can only see a single trap event and a corresponding transition from a 1/f spectrum to a Lorentzian. Many researchers consider traps in the oxide layer to be the origin of RTSs. The observation of a RTS generated by a single trap suggests that a RTS is the microscopic source of 1/f noise. Based on this observation, 1/f noise is explained as a superposition of individual trapping events; traps distributed over the oxide layer lead to a wide distribution of time constants [4].

In the time domain, besides the switching amplitude, a RTS is defined by the up-and-down time constants, which are attributed to an occupied and an empty state, respectively. The up-and-down times have an exponential distribution strongly dependent on temperature [7, 9]. Machlup investigated the current noise caused by a single trap [12]. As for generation-recombination (g-r) noise, the spectrum of RTS noise also has a Lorentzian shape. It is an open question whether this similarity is only formal because a g-r noise signal is Gaussian, whereas a RTS is merely in an up-and-down state [5].

This paper investigates the g-r noise caused by a single trap of a semiconductor material. We start with the well-known master-equation approach providing g-r bulk noise in terms of parameters of conduction electrons. We show that g-r bulk noise can also be attributed to a random succession of elementary g-r pulses generated by all traps. This enables g-r bulk noise to be formulated in terms of the numbers of traps applying also to just one single active trap. The g-r noise produced by such a single active trap is equivalent to Machlup's noise spectrum. We also discuss a possible overlap of succeeding g-r pulses.

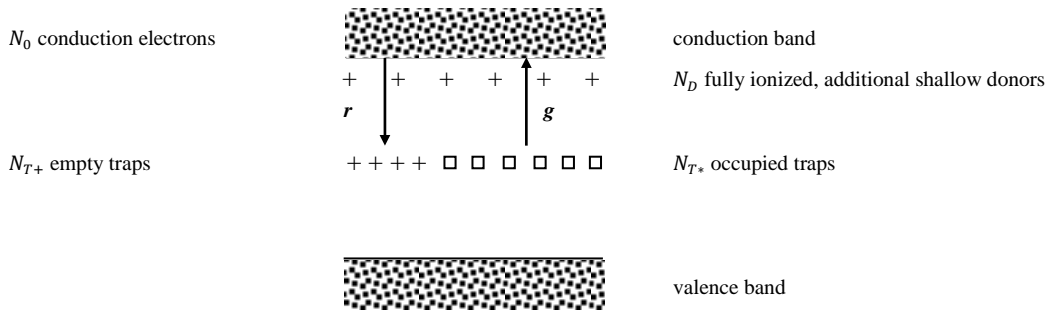

Fig. 1. Electron transitions for an extrinsic n-type semiconductor with traps. Under steady-state conditions, $N_{T+}$ is the number of empty traps and $N_{T*}$ the number of occupied traps. $N_T = N_{T+} + N_{T*}$ is the number of traps which are neutral when occupied. $N_0$ is the number of conduction electrons and $N_D$ the number of fully ionized, additional shallow donors.

## 2. Fluctuations in an extrinsic semiconductor with traps

In a doped n-type semiconductor, we only consider transitions between the level of traps and the conduction band to produce fluctuations (see Fig. 1). Thermal noise that is always present will be neglected. For steady-state conditions, $g_0$ and $r_0$ are the generation and recombination probabilities per time unit, respectively, and $N_0$ the number of conduction electrons. For $N_0 \gg 1$, the fluctuations of conduction electrons tend to be normally distributed about $N_0$ with $\overline{\Delta N_{gr}^2}$ being the mean square fluctuations of conduction electrons contributing to the g-r process. Under these conditions, the master-equation approach leads to [13-15]

$$g_0 = \overline{\Delta N_{gr}^2}/\tau_{gr} \tag{1}$$

and

$$S_{gr}(f) = 4\left(\frac{I_0}{N_0}\right)^2 \frac{g_0 \tau_{gr}^2}{1+(2\pi f \tau_{gr})^2}. \tag{2}$$

$S_{gr}(f)$ is the power-spectral density of current fluctuations due to an applied current $I_0$ versus frequency $f$ and $\tau_{gr}$ is the g-r relaxation time. Let the semiconductor volume be 1cm³ so that the number of electrons and traps coincides with their concentration. The number of fully-ionized, additional shallow donors is $N_D$ and the number of traps in the semiconductor material, $N_T$. For steady-state conditions, the mean number of conduction electrons is

$$N_0 = N_{T+} + N_D \tag{3}$$

and

$$g_0 = \gamma N_{T*} \quad \text{and} \quad r_0 = \rho N_0 N_{T+} \tag{4}$$

with $\gamma$ and $\rho$ being the coefficients of generation and recombination, respectively. Here, the mean number of occupied traps is $N_{T*}$ and of empty traps, $N_{T+} = N_T - N_{T*}$. The master-equation approach provides the g-r relaxation time by [13-15]

$$\frac{1}{\tau_{gr}} = \gamma + \rho N_0 + \rho N_{T+}. \tag{5}$$

Expressing the numbers of occupied/empty traps and of conduction electrons in relation to their corresponding relaxation-time constants yields

$$g_0 = N_{T*}/\tau_{T*} \quad \text{and} \quad r_0 = N_{T+}/\tau_{T+} = N_0/\tau_0. \tag{6}$$

By comparison with Eq. (4), we find

$$\gamma = 1/\tau_{T*} \quad \text{and} \quad \rho N_0 = 1/\tau_{T+} \quad \text{and} \quad \rho N_{T+} = 1/\tau_0. \tag{7}$$

Substituting this into Eq. (5), the g-r relaxation time can be expressed by

$$\frac{1}{\tau_{gr}} = \frac{1}{\tau_{T*}} + \frac{1}{\tau_{T+}} + \frac{1}{\tau_0}. \tag{8}$$

A large number of additional shallow donors $N_D \gg N_T$ reduces this equation to

$$\frac{1}{\tau_{gr}} \approx \frac{1}{\tau_{T*}} + \frac{1}{\tau_{T+}}. \tag{9}$$

Using Eq. (1), the mean square fluctuations of conduction electrons are obtained by

$$\frac{1}{\overline{\Delta N_{gr}^2}} = \frac{1}{N_{T*}} + \frac{1}{N_{T+}} + \frac{1}{N_0}. \tag{10}$$

### 2.1. *An alternative definition of the generation rate*

By detailed balance ($g_0 = r_0$) Eq. (6) leads to

$$\frac{\tau_{T+}}{\tau_{T*}} = \frac{N_T - N_{T*}}{N_{T*}} = \frac{1-f_T}{f_T}. \tag{11}$$

Herein, $f_T$ is the occupancy of the traps by an electron defined by

$$f_T = \frac{N_{T*}}{N_T} = \frac{\tau_{T*}}{\tau_T}, \tag{12}$$

where

$$\tau_T = \tau_{T+} + \tau_{T*} \tag{13}$$

is the time between successive generations at a trap. Combining Eq. (6) with Eq. (11), an alternative definition of the generation rate is found by

$$g_0 = \frac{N_T}{\tau_T} \tag{14}$$

relating the generation rate to the parameters of traps rather than to the parameters of conduction electrons as is defined in Eq. (1).

### 2.2. *Fluctuations due to an individual trap*

Consequently, the generation rate due to an individual trap is given by

$$\frac{1}{\tau_T} = \frac{g_0}{N_T}. \qquad (15)$$

After generation, an electron remains in the conduction band for the g-r relaxation time $\tau_{gr}$ before it recombines into an arbitrary empty trap. This gives rise to a random succession of elementary g-r pulses $h_{gr}(t)$. This pulse train can be described by

$$i_{gr}(t) = \sum_{k=-\infty}^{+\infty} h_{gr}(t - \vartheta_k), \qquad (16)$$

$\vartheta_k$ being the occurrence time of the k-th event. Fig. 2 shows the current fluctuations $i_{gr}(t)$ due to an individual trap. It is emphasized that the g-r pulse train seen in this figure is <u>not</u> a RTS, but a shot-noise signal including a possible overlap of succeeding g-r pulses. Such an overlap, as is illustrated in Fig. 2 for the third and fourth g-r pulse, will be investigated in section 3.3. Equating Eqs. (1) and (14) and making use of Eq. (10), we find

$$\frac{\tau_{gr}}{\tau_T} = \frac{\overline{\Delta N_{gr}^2}}{N_T} = f_T(1 - f_T)\frac{(1-f_T)+N_D/N_T}{(1+f_T)(1-f_T)+N_D/N_T}. \qquad (17)$$

It is easily seen that $0 < \tau_{gr}/\tau_T \leq 1/4$.

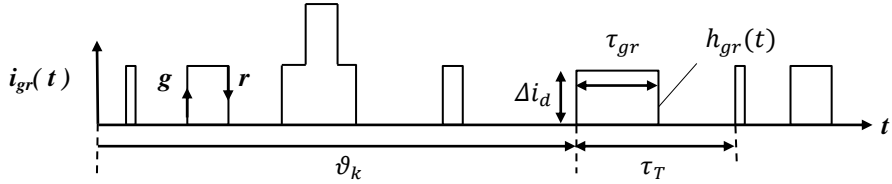

Fig. 2. Current fluctuations $i_{gr}(t)$ of an individual trap consisting of a random succession of g-r pulses $h_{gr}(t)$ with switching amplitude $\Delta i_d$ and g-r relaxation time $\tau_{gr}$. The time between succeeding generations is $\tau_T$. A mean drift current is $\Delta i_d = e\,v_d/L$ where $e$ is the elementary charge and $L$ the length of the sample. A mean drift velocity is $v_d = \mu E_0$ where $\mu$ is the mobility and $E_0$ an applied electric field.

### 2.3. *The g-r shot noise in the bulk*

Correspondingly, the g-r current fluctuations originating from all of the $N_T$ traps in the bulk are represented by

$$I_{gr}(t) = \sum_{k=-\infty}^{+\infty} h_{gr}(t - \theta_k), \qquad (18)$$

where g-r pulses occur at a generation rate of $g_0 = N_T/\tau_T$. The random succession of elementary pulses is generally referred to as shot noise [16-17]. We introduce the term[1] "g-r shot noise" to characterize the random succession of g-r pulses. Applying Carson's theorem for elementary events [18], the power spectral density of $I_{gr}(t)$ is obtained by

$$S_{gr}^{shot}(f) = 2g_0\,\overline{|H_{gr}(f)|^2} \qquad (19)$$

disregarding a DC (Direct Current) component, which is discussed in the next section. $H_{gr}(f)$ is the Fourier transform of $h_{gr}(t)$. For the rectangular current pulse with switching amplitude $\Delta i_d$ and exponentially distributed g-r relaxation time, we obtain

$$\overline{|H_{gr}(f)|^2} = \frac{2\,\Delta i_d^2 \tau_{gr}^2}{1+(2\pi f \tau_{gr})^2}. \qquad (20)$$

Considering that a mean current

$$I_0 = N_0\,\Delta i_d, \qquad (21)$$

the g-r shot noise in the bulk can be expressed as

$$S_{gr}^{shot}(f) = 4\left(\frac{I_0}{N_0}\right)^2 \frac{g_0 \tau_{gr}^2}{1+(2\pi f \tau_{gr})^2} \qquad (22)$$

in agreement with Eq. (2). Succeeding g-r pulses originating from different traps strongly overlap. This leads to a quasi-continuous random function $I_{gr}(t)$ with a normal probability distribution (see Appendix A). This is the basic condition of the master-equation approach for deriving g-r noise. The g-r shot-noise interpretation is fully consistent with the master-equation approach to g-r noise.

---

[1] Carson's theorem for elementary events (see Eq. (19)) is used to derive the power spectral density of both g-r shot noise in a semiconductor and of shot noise in a vacuum tube operating in the saturation region. However, their power spectral densities differ due to different physical mechanisms: for g-r shot noise only a small fraction of an elementary charge $e$ is transferred per pulse which leads to $S_{gr}^{shot} \propto I_0^2$ (see Eq. (22)). In a vacuum tube, a full elementary charge $e$ is transferred per pulse resulting in $S_{tube}^{shot} = 2eI_0$ (see [16-17]).

Some additional information not provided by the master-equation approach is obtained by the g-r shot-noise interpretation: 1) applying Campbell's theorem, the mean of $I_{gr}(t)$ can be calculated, and 2) the g-r shot-noise interpretation applies not only for g-r bulk noise, but also for a single trap. This is shown in the next sections.

### 2.4. *The fraction of conduction electrons contributing to the g-r process*

Applying Campbell's theorem [16, 19], the mean of $I_{gr}(t)$ is obtained by (see Appendix A)

$$\overline{I_{gr}} = g_0 \tau_{gr} \Delta i_d = \frac{\overline{\Delta N_{gr}^2}}{N_0} I_0 \qquad (23)$$

relating the mean of g-r noise current $\overline{I_{gr}}$ to an applied current $I_0$. Introducing

$$\eta_{gr} \equiv \frac{\overline{\Delta N_{gr}^2}}{N_0} \qquad (24)$$

and applying Eq. (10) leads to

$$\eta_{gr} = \frac{f_T(1-f_T)}{(1+f_T)(1-f_T)+N_D/N_T}. \qquad (25)$$

It is easily verified that $0 < \eta_{gr} < \frac{1}{2}$.

$$\overline{I_{gr}} = \eta_{gr} I_0 \qquad (26)$$

is the mean current, and

$$\overline{N_{gr}} = \eta_{gr} N_0 \qquad (27)$$

is the mean number of conduction electrons contributing to the g-r process.

$$\eta_{gr} = \frac{\overline{N_{gr}}}{N_0} \qquad (28)$$

can be interpreted as the fraction of the conduction electrons contributing to the g-r process; $\eta_{gr}$ may also be called the efficiency of the g-r process. Comparing Eqs. (24) and (28) leads to

$$\overline{\Delta N_{gr}^2} = \overline{N_{gr}} \qquad (29)$$

showing that the g-r number fluctuations obey Poisson-statistics[2], i.e. the generation of g-r pulses occurs at random. Consequently, the generation rate in Eq. (1) can also be expressed as

$$g_0 = \overline{N_{gr}}/\tau_{gr}. \qquad (30)$$

Equating this with Eq. (14), we obtain

$$\overline{N_{gr}} = N_T \frac{\tau_{gr}}{\tau_T} \qquad (31)$$

relating the number of the conduction electrons contributing to the g-r process to the number of traps.

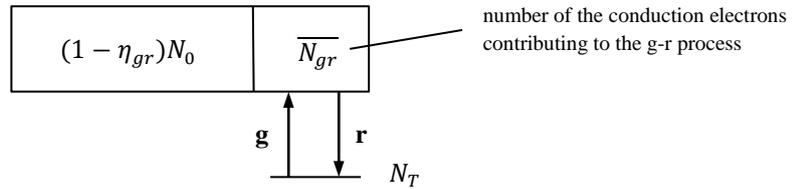

Fig. 3. Number of the conduction electrons contributing to the g-r process (right reservoir) in comparison to the number of conduction electrons which do not contribute to the g-r process (left reservoir). The electrons are not locally concentrated in the right or left reservoir; this figure merely illustrates the fraction of the corresponding contributions.

### 2.5. *Amplitude distribution of total current fluctuations*

According to Eq. (27), the number of conduction electrons which do not contribute to the g-r process is given by $(1 - \eta_{gr})N_0$ and exhibits no fluctuations. The corresponding contributions are illustrated in Fig. 3. Due to jumps between electron states within the conduction band, electrons may jump between the left and the right reservoirs. Under the supposition that these jumps are much faster than the g-r relaxation time, these jumps do not affect the noise behavior of the g-r process. Under this condition, the total current fluctuations are given by

$$I_{tot}(t) = (1 - \eta_{gr}) I_0 + I_{gr}(t), \qquad (32)$$

leading to

$$\overline{I_{tot}} = I_0 \qquad (33)$$

and

$$\overline{\Delta I_{tot}^2} = \overline{\Delta I_{gr}^2}. \qquad (34)$$

---

[2] Like g-r shot noise in a semiconductor also shot noise in a vacuum tube operating in the saturation region can be modeled by a Poisson process [16-17].

The total current is carried by all conduction electrons. The total current fluctuations, however, are exclusively caused by the variance of the conduction electrons contributing to the g-r process. Fig. 4 illustrates the corresponding contributions.

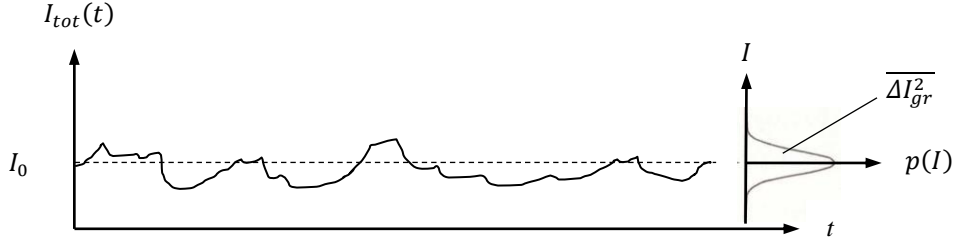

Fig. 4. Amplitude distribution of total current fluctuations $I_{tot}(t)$ with probability distribution $p(I)$.

## 3. The transition from g-r noise in the bulk to g-r noise due to a single active trap

Eq. (22) expresses the g-r shot noise in the bulk as

$$S_{gr}^{shot}(f) = 4I_0^2 \frac{\eta_{gr}^2}{g_0} \frac{1}{1+(2\pi f \tau_{gr})^2}, \qquad (35)$$

emphasizing that only the fraction $\eta_{gr}$ of the conduction electrons contributes to the g-r process. Substituting $g_0 = N_T/\tau_T$ leads to

$$S_{gr}^{shot}(f) = 4I_0^2 \frac{\eta_{gr}^2}{N_T} \frac{\tau_T}{1+(2\pi f \tau_{gr})^2}, \qquad (36)$$

which interprets the g-r shot noise in terms of the number of traps. This relation applies for an arbitrary number of traps and – with some modifications – also for $N_T = 1$, i.e. for a single active trap. By reducing the number of traps from $N_T \gg 1$ to $N_T = 1$, the amplitude distribution of the g-r noise signal changes from a continuous normal distribution seen in Fig. 4 into a discrete skewed distribution discussed in section 3.4. Eqs. (2) and (22) apply for a normal amplitude distribution; the range of validity is extended by Eq. (36) which also applies for discrete skewed distributions emerging for low numbers of traps.

The g-r shot-noise interpretation suggests that the random succession of g-r pulses observed in MOSFETs and other semiconductors is due to a single active trap ($N_T = 1$). We exclude possible additional, shallow donors contributing to the conduction mechanism ($N_D = 0$). This is, however, an exceptional case. All other cases – like $N_T \geq 1$ and $N_D \geq 1$ but also $N_T \geq 2$ and $N_D \geq 0$ – can be handled within the scope of g-r shot noise in the bulk (see section 2).

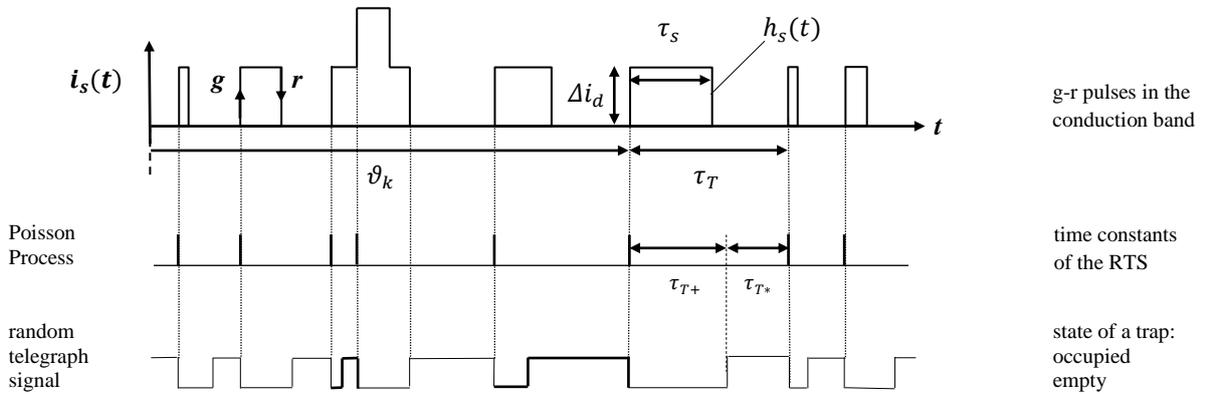

Fig. 5. Top: Current fluctuations $i_s(t)$ of a single trap represented by a random succession of g-r pulses $h_s(t)$ with switching amplitude $\Delta i_d$ and g-r relaxation time $\tau_s$. Middle: The underlying spike train is a Poisson process; each spike triggers an elementary g-r pulse $h_s(t)$. The time between successive generations is $\tau_T = \tau_{T+} + \tau_{T*}$. Bottom: Random telegraph signal representing an occupied or an empty state of a single trap. The bold pulses of the RTS indicate occupation of the empty single trap by an electron transferred from a neighboring trap.

## 3.1. *Machlup's noise spectrum for a single active trap*

The master-equation approach does not apply for a single active trap, i.e. for $N_T = 1$ and $N_D = 0$. We can take advantage of an alternative approach to the noise spectrum provided by Machlup [12]. In the following, we compare Machlup's noise spectrum with the g-r shot noise due to a single trap.

According to Machlup's approach [12], we regard a single trap being either occupied by an electron or empty. The average lifetime of the occupied state is $\tau_{T*}$ and that of the empty state, $\tau_{T+}$ (see the middle of Fig. 5). The corresponding RTS is shown on the bottom of Fig. 5. Mathematically an (asymmetric) RTS is described by a two-state Markov process [20-22]. Machlup derived the spectral density (dimension 1/Hz) of a RTS by

$$\tilde{S}_{RTS}(f) = \frac{4}{(\tau_{T*}+\tau_{T+})\left[\left(\frac{1}{\tau_{T*}}+\frac{1}{\tau_{T+}}\right)^2 + (2\pi f)^2\right]}. \tag{37}$$

This gives rise to a random succession of g-r pulses in the conduction band (see the top of Fig. 5). Multiplication with the switching amplitude squared yields the power spectral density (dimension A²/Hz) of this g-r pulse train by

$$S_M(f) = \Delta i_d^2 \, \tilde{S}_{RTS}(f). \tag{38}$$

The subscript *"M"* stands for Machlup. For convenience we introduce the relaxation time of the RTS by

$$\frac{1}{\tau_{RTS}} \equiv \frac{1}{\tau_{T*}} + \frac{1}{\tau_{T+}}. \tag{39}$$

Comparing this equation with Eqs. (8) and (9) suggests that $\tau_{RTS}$ can also be interpreted as the g-r relaxation time of conduction electrons. This is shown for the following: for a single trap Eq. (4) reduces to

$$g_0 = \gamma N_{T*} \quad \text{and} \quad r_0 = \rho N_{T+}. \tag{40}$$

This leads to the single trap's g-r relaxation time by

$$\frac{1}{\tau_s} \equiv \gamma + \rho. \tag{41}$$

Obviously, $\tau_s \geq \tau_{gr}$ i.e. $\tau_s$ is larger than or equal to the g-r relaxation time in the bulk. Comparing Eq. (40) with Eq. (6) yields $\gamma = 1/\tau_{T*}$ and $\rho = 1/\tau_{T+}$ resulting in

$$\tau_{RTS} = \tau_s. \tag{42}$$

In words, the relaxation time of the RTS is identical with the g-r relaxation time due to a single trap[3]. Considering $\tau_T = \tau_{T*} + \tau_{T+}$ (see Fig. 5), Machlup's noise spectrum is transformed into

$$S_M(f) = 4 \, \Delta i_d^2 \, \frac{\tau_s^2}{\tau_T} \, \frac{1}{1+(2\pi f \tau_s)^2} \tag{43}$$

depending on the parameters of a g-r pulse. Applying Eqs. (12) and (41) leads to

$$\frac{\tau_s}{\tau_T} = f_T(1 - f_T). \tag{44}$$

which is shown in Fig. 6.a versus trap's occupancy. Substitution into Eq. (43) results in

$$S_M(f) = 4 f_T (1 - f_T) \, \Delta i_d^2 \, \frac{\tau_s}{1+(2\pi f \tau_s)^2}. \tag{45}$$

This form of Machlup's noise spectrum has been applied by several authors [1, 7].

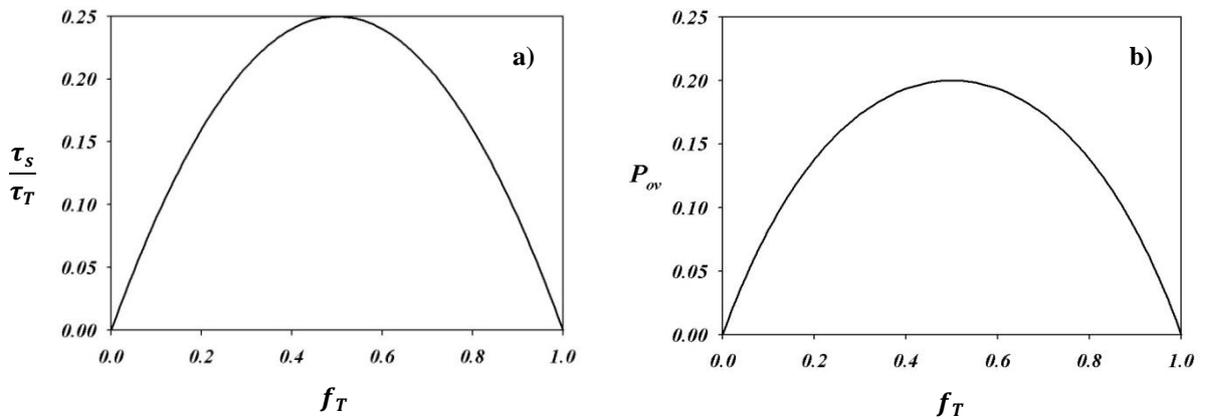

Fig. 6.a) The quotient $\tau_s/\tau_T$ and b) the probability of an overlap $P_{ov}$ versus trap's occupancy $f_T$.

---

[3] Machlup equates the g-r relaxation time with the lifetime of an empty trap, i.e. he defines $\tau_s \equiv \tau_{T+}$. Most researchers [1-9] agree with Machlup's interpretation. This interpretation however is inconsistent with above definitions of $\tau_s$ and $\tau_{RTS}$.

### 3.2. *The g-r shot noise resulting from a single active trap*

According to Eq. (14), for a single active trap ($N_T = 1$) the rate of generation is given by $g_0 = 1/\tau_T$. This leads to a g-r pulse train $i_s(t)$ seen at the top of Fig. 5. Applying Carson's theorem, the power-spectral density of this g-r pulse train is

$$S_{trap}^{single}(f) = \frac{2}{\tau_T} \overline{|H_s(f)|^2} \tag{46}$$

aside from a DC component, which is investigated below. Herein

$$H_s(f) = \int_{-\infty}^{\infty} h_s(t) \exp(-i2\pi f t)\, dt \tag{47}$$

is the Fourier transform of $h_s(t)$. For a rectangular current pulse $h_s(t)$ with switching amplitude $\Delta i_d$ and lifetime $\tau_h$, we obtain

$$H_s(f) = \Delta i_d \tau_h \frac{\sin(\pi f \tau_h)}{\pi f \tau_h}. \tag{48}$$

Machlup defined g-r shot noise as a succession of g-r pulses with <u>identical</u> lifetimes $\tau_h$ leading to

$$S_h(f) = \frac{2}{\tau_T} |H_s(f)|^2 = \frac{2}{\tau_T} \left( \Delta i_d \tau_h \frac{\sin(\pi f \tau_h)}{\pi f \tau_h} \right)^2. \tag{49}$$

However, the lifetime of conduction electrons (= g-r relaxation time $\tau_s$) is exponentially distributed with the probability density function

$$t \geq 0: \qquad p_{\tau_s}(t) = \frac{1}{\tau_s} \exp\left(-\frac{t}{\tau_s}\right) \tag{50}$$

yielding

$$\overline{|H_s(f)|^2} = 2 \left|\overline{H_s(f)}\right|^2 = \frac{2\, \Delta i_d^2\, \tau_s^2}{1+(2\pi f \tau_s)^2}. \tag{51}$$

Substituting this into Eq. (46) leads to the power spectral density due to a single active trap by

$$S_{trap}^{single}(f) = \overline{S_h(f)} = 4\, \Delta i_d^2 \frac{\tau_s^2}{\tau_T} \frac{1}{1+(2\pi f \tau_s)^2}. \tag{52}$$

A comparison with Eq. (43) reveals that[4]

$$S_{trap}^{single}(f) = S_M(f) \tag{53}$$

demonstrating that Machlup's noise spectrum is equal to the g-r shot noise of a single trap[5]. Applying Campbell's theorem [16, 19] the mean of $i_s(t)$ is obtained by

$$\overline{i_s} = g_0 \overline{\int_{-\infty}^{\infty} h_s(t)\, dt} = g_0 \overline{H_s(0)}. \tag{54}$$

Using Eq. (51) leads to

$$I_0 = \overline{i_s} = g_0 \tau_s\, \Delta i_d = \frac{\tau_s}{\tau_T} \Delta i_d \tag{55}$$

relating the DC component $I_0$ to the switching amplitude $\Delta i_d$ of the g-r pulse and transforms Eq. (52) into

$$S_{trap}^{single}(f) = 4\, I_0^2 \frac{\tau_T}{1+(2\pi f \tau_s)^2}. \tag{56}$$

In this form, spectral estimates of the time series of a single trap $i_s(t)$ provide the parameters $\tau_T$ and $\tau_s$. For an example, see Fig. 2 in reference [7].

Comparing Eq. (55) with Eq. (21), we find the mean number of conduction electrons by $N_0 = \tau_s/\tau_T$ being no longer an integer like in Eq. (3) but a real number ≤ ¼ (see Fig. 6.a). By analogy to Eq. (31), the mean number of conduction electrons contributing to the g-r process is $\overline{N_{gr}} = g_0 \tau_s = \tau_s/\tau_T$. Hence, in combination with Eq. (44) we find

$$N_0 = \overline{N_{gr}} = \overline{\Delta N_{gr}^2} = f_T(1-f_T) = \tau_s/\tau_T. \tag{57}$$

By analogy to Eq. (28), for a single trap the fraction of the conduction electrons contributing to the g-r process is

$$\eta_s = \overline{N_{gr}}/N_0 = 1, \tag{58}$$

i.e. each conduction electron fully contributes to the g-r process. This is in strong contrast to g-r bulk noise, where only a fraction $\eta_{gr} < ½$ of the conduction electrons contributes to the g-r process. The transition from g-r bulk noise (see Eq. (36)) to discrete switching due to a single active trap (see Eq. (56)) is accomplished by the following replacements:

$$N_T \to 1 \quad \text{and} \quad \eta_{gr} \to \eta_s \quad \text{and} \quad \tau_{gr} \to \tau_s. \tag{59}$$

---

[4] Machlup misses this identity because he defines g-r shot noise as a succession of g-r pulses with <u>identical</u> relaxation times (see Eq. (49)).
[5] The power-spectral density of a (symmetrical) RTS was first derived by Rice [23]; it exhibits a Lorentzian shape. Rice mentions that this type of power spectra is obtained for shot noise.

### 3.3. *Overlap of succeeding g-r pulses*

If the g-r relaxation time of an elementary g-r pulse is longer than that of the empty single trap (e.g. the fifth g-r pulse in Fig. 5), the electron in the conduction band will have no empty trap in which to recombine. This suggests that an empty single trap is occupied by an electron transferred from a neighboring trap (as is demonstrated by bold RTS pulses in Fig. 5). The electron in the conduction band may recombine into the empty neighboring trap. The same mechanism may also lead to overlapping g-r pulses (see third and fourth g-r pulse in Fig. 5). Two or more neighboring traps may also be involved[6], as is the case for multiple overlapping of g-r pulses which are discussed in this and the next section.

The random succession of g-r pulses seen at the top of Fig. 5 is g-r shot noise. In contrast to a RTS, succeeding g-r pulses can overlap (see the third and fourth g-r pulse in Fig. 5); this occurs, when $\tau_T \leq \tau_s$. Both $\tau_T$ and $\tau_s$ are statistical variables. We therefore determine the probability that $\tau_T \leq \tau_s$ in two steps. In a first step, we calculate the probability that $\tau_T \leq \tau_h$ for identical g-r lifetime $\tau_h$, which is averaged over $\tau_s$ in a second step. The times between succeeding generations are exponentially distributed with the probability-density function $p_{\tau_T}(t)$. The probability for an overlap of succeeding g-r pulses with identical g-r lifetime $\tau_h$ is

$$\int_0^{\tau_h} p_{\tau_T}(t)\, dt = \int_0^{\tau_h} \frac{1}{\tau_T} exp\left(-\frac{t}{\tau_T}\right) dt = 1 - exp\left(-\frac{\tau_h}{\tau_T}\right). \tag{60}$$

The average over $\tau_h$ yields the probability of an overlap by[7]

$$P_{ov} \equiv prob\{\tau_T \leq \tau_s\} = \int_0^\infty \left[1 - exp\left(-\frac{t}{\tau_T}\right)\right] p_{\tau_s}(t)\, dt = \frac{\frac{\tau_s}{\tau_T}}{1 + \frac{\tau_s}{\tau_T}}. \tag{61}$$

Simulating a random succession of $10^6$ g-r pulses, the probability $P_{ov}$ has been estimated for several values of $\tau_s/\tau_T \leq 0.25$; the agreement with calculated values is excellent (see Table 1).

Table 1. The calculated and simulated probability of an overlap $P_{ov}$ for several values of $\tau_s/\tau_T$. Averages for $10^6$ samples.

| $\tau_s/\tau_T$ | 0.01 | 0.05 | 0.10 | 0.15 | 0.167 | 0.20 | 0.25 |
|---|---|---|---|---|---|---|---|
| $P_{ov}$ calculated | 0.010 | 0.048 | 0.091 | 0.130 | 0.143 | 0.167 | 0.200 |
| $P_{ov}$ simulated | 0.0099 | 0.047 | 0.091 | 0.130 | 0.143 | 0.166 | 0.200 |

Substituting Eq. (44) yields

$$P_{ov} = \frac{f_T(1-f_T)}{1 + f_T(1-f_T)}. \tag{62}$$

Fig. 6.b shows the probability of an overlap $P_{ov}$ versus trap's occupancy $f_T$. For low and high occupancy, $P_{ov} \ll 1$, i.e., an overlap is rather improbable. The maximum of $P_{ov} = 1/5$ is found at $f_T = 0.5$; $P_{ov} = 1/5$ implies that, on the average, every 5th g-r pulse overlaps with the succeeding one.

### 3.4. *Simulated g-r pulse train due to a single active trap*

Figure 7 shows simulated g-r pulse trains for several values of $\tau_s/\tau_T$. The switching amplitude is $\Delta i_d = 1$. The mean time between successive generations $\tau_T$ is the same for all four plots. Increasing the g-r relaxation time $\tau_s$ leads to an increasing overlap of succeeding g-r pulses, including multiple overlaps. The Fortran program for the simulation procedure can be found in the Appendix C.

The simulated g-r pulse trains enable a qualitative estimate of the amplitude distribution of current fluctuations $i_s(t)$. There is high probability for state 0 and a decreasing probability for states 1, 2, and 3 leading to a skewed discrete amplitude distribution.

For $\tau_s/\tau_T \leq 0.01$, i.e. for low and high occupancy, the g-r pulse train seems to be exclusively in a down-state 0 and an up-state 1; it looks like a RTS, but isn't! The probability for an overlap $P_{ov} \approx 0.01$, i.e. on the average every 100th pulse exhibits an overlap. The sixteen g-r pulses in Fig. 7 shown for $\tau_s/\tau_T = 0.01$ exhibit no overlap; extending the length of this g-r pulse train should lead to overlapping g-r pulses.

---

[6] We ignore possible interactions between traps like Coulomb interactions as have been discussed by Kirton and Uren [4].
[7] Corresponding considerations for $P_{ov}$ in the bulk are presented in Appendix B.

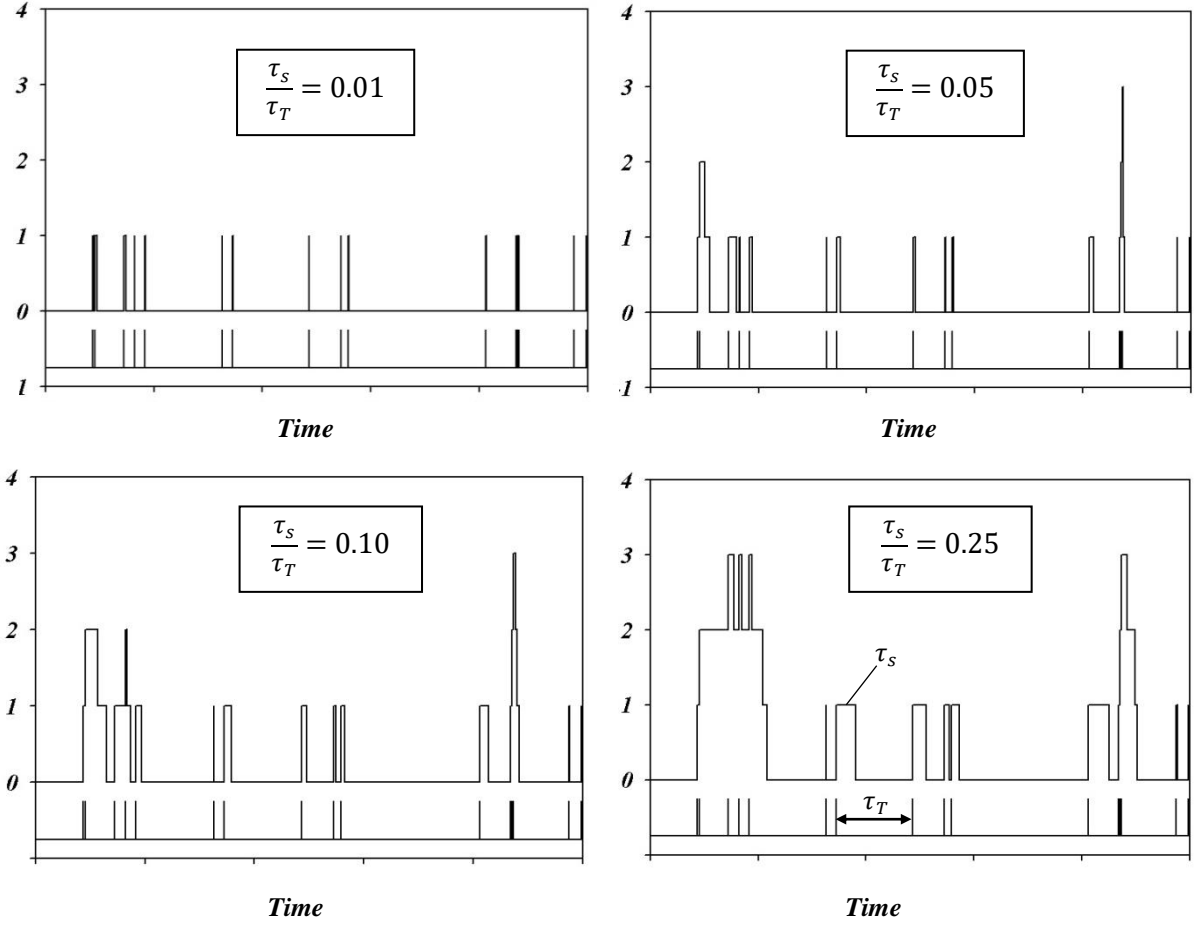

Fig. 7. Simulated g-r pulse trains $i_s(t)$ due to a single trap for several values of $\tau_s/\tau_T$. The baseline for all four plots is a Poisson process with $\tau_T$ being the time between successive generations. The spikes indicate the generation of an electron giving rise to a g-r pulse with $\tau_s$ being the g-r relaxation time due to a single trap. The switching amplitude is $\Delta i_d = 1$.

### 3.5. *Distribution of up-and-down times*

A defining sketch of the up-and-down times is shown in Fig. 8. The up time $\tau_{up}$ is exponentially distributed; it is identical with the relaxation time of a g-r pulse

$$\tau_{up} \equiv \tau_s. \tag{63}$$

The down time $\tau_{down}$ is defined as the time between the recombination of an electron and the start of the next g-r pulse. The down time is observed only if $\tau_{up} < \tau_T$; otherwise succeeding g-r pulses overlap. As an example see the overlap of the third and fourth g-r pulse in Fig. 5 where $\tau_s < \tau_T$. Such overlapping g-r pulses (including multiple overlaps) have to be omitted for the statistical analysis of the down times.

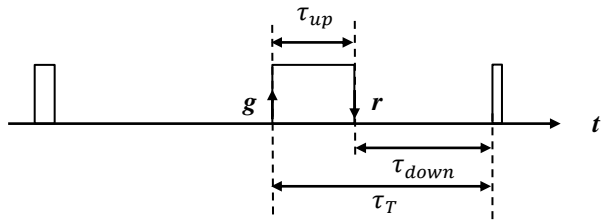

Fig. 8. Definition sketch of up-and-down times. The up time is identical to the g-r relaxation time $\tau_s$ of a single trap.

We simulated a g-r pulse train containing $10^5$ g-r pulses. Excluding overlapping g-r pulses, the distribution of the up-and-down times was estimated. As is seen in Fig. 9, the up time is exponentially distributed with mean value $\tau_s$. For large $t$, the down time approaches an exponential distribution with mean value $\tau_T$; for small $t$, the down time deviates from the exponential decay.

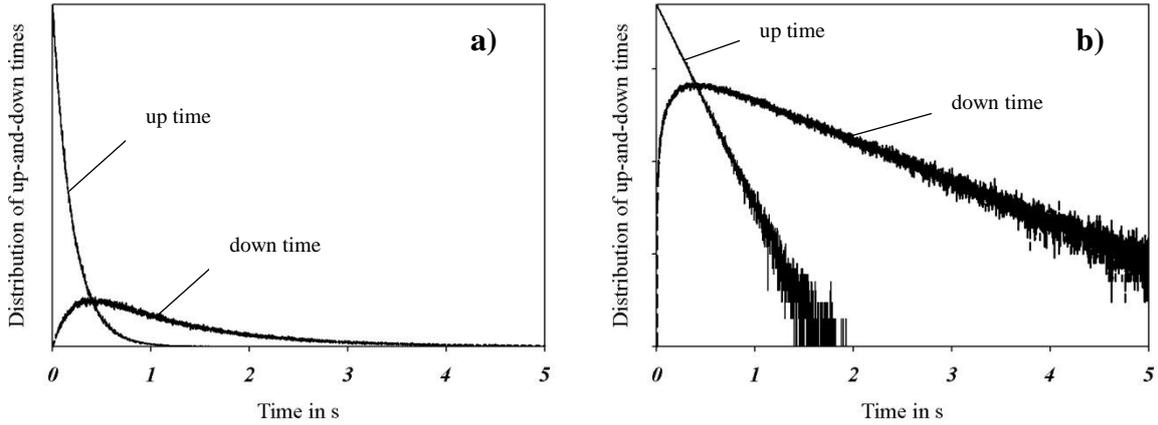

Fig. 9. Distribution of up-and-down times excluding overlapping g-r pulses. Estimates for $10^5$ samples in a) lin-lin scale and b) log-lin scale. $\tau_{up} = \tau_s = 0.25\ s$ and $\tau_T = 1\ s$.

## 4. Results and Discussion

This paper presents the transition of g-r noise in bulk semiconductors to discrete switching in small-area semiconductors. We apply an alternative approach to g-r noise in a bulk semiconductor by describing the g-r noise as a random succession of elementary g-r pulses. Such a g-r pulse train is referred to as "g-r shot noise". We show that the g-r shot-noise interpretation is fully consistent with the master-equation approach to g-r noise. In contrast to the master-equation approach, the g-r shot-noise interpretation makes it possible to formulate g-r bulk noise in terms of the number of traps. Discrete switching is obtained by reducing the number of traps to just one single active trap.

This suggests the following interpretation: a RTS characterizing the occupancy of a single trap can be described by a two-state Markov process. In the conduction band, this gives rise to a random succession of rectangular g-r pulses. We show that the relaxation time of the two-state Markov process is identical with the g-r relaxation time. Consequently, Machlup's noise spectrum is equal to the g-r shot noise due to a single trap. An overlap of succeeding g-r pulses is explained by an empty single trap which is occupied by an electron transferred from a neighboring trap. The electron in the conduction band may then recombine to an empty neighboring trap. Multiple overlap of g-r pulses suggests that two or more neighboring traps are involved.

Our interpretation does not coincide with the interpretation suggested by several authors [1-9]. Following Machlup [12], the up-and-down time constants of a g-r pulse train are usually interpreted as the lifetime of the empty and occupied state of a single trap in an oxide layer. This excludes an overlap of succeeding g-r pulses; overlapping pulses are attributed to emissions of two or more active single traps.

In contrast to this interpretation, the g-r shot-noise interpretation may shed a new light on overlapping g-r pulses. We calculated the probability of an overlap of succeeding pulses and showed that this probability depends on trap's occupancy. An overlap is improbable for low and high occupancy of traps. For intermediate occupancy, the probability of an overlap can be up to 1/5, i.e. on the average up to every 5[th] g-r pulse can overlap with the succeeding one. Simulating g-r shot noise we showed that an overlap of succeeding g-r pulses leads to a wide variety of patterns similar to those observed in a MOSFET [4]. The agreement between empirical and simulated signals is open to question and must be checked by an appropriate statistical analysis.

Excluding overlapping g-r pulses, we estimated the up-and-down times of a simulated g-r pulse train. In agreement with empirical findings, the up times are found exponentially distributed. For large times, the down times rapidly approach an exponential distribution; however, the down times deviate from the exponential decay for small times. Kirton and Uren also report a departure from an exponential decay for small times; they attribute this departure to a limited resolution of sampling time. Our simulations suggest that this departure is systematic and not due to a limited resolution.

We suggest that the origin of g-r pulse trains in a device like a MOSFET is due to a single trap in the semiconductor material. Alternatively, such a single trap can also be located in an oxide layer. Kirton and Uren also assume traps in the oxide layer to be an origin of RTSs in small-area MOSFETs. They explain 1/f noise in the bulk by a superposition of individual trapping events with a wide distribution of time constants. We do not agree with this interpretation: 1/f noise in a bulk semiconductor can also be explained by an intermittent g-r process [24, 25], where the generation process is gated by on-off states. The intermittent g-r process does not need traps in an oxide layer exhibiting a wide distribution of time constants.

## Appendix A. Approach of distribution of $I_{gr}(t)$ to a normal law

According to Eq. (18), the current fluctuations for g-r shot noise in the bulk are given by

$$I_{gr}(t) = \sum_{k=-\infty}^{+\infty} h_{gr}(t - \vartheta_k), \qquad (A.1)$$

where g-r pulses occur at a generation rate $g_0$. As was shown by Rice [23], a normal distribution of $I_{gr}(t)$ is approached in the following way: the n-th semi-invariant of the probability distribution of $p(I)$ is

$$\lambda_n = g_0 \int_{-\infty}^{\infty} h_{gr}^n(t)\,dt = g_0 \tau_{gr}\, \Delta i_d^n. \qquad (A.2)$$

It can be shown that $\lambda_1 = \overline{I_{gr}}$ and $\lambda_2 = \overline{\Delta I_{gr}^2}$. Hence, when we set $n = 1$ and $2$, Eq. (A.2) yields Campbell's theorem [16, 19]. Applying Eq. (30), we find

$$\overline{I_{gr}} = \overline{N_{gr}}\, \Delta i_d \qquad (A.3)$$

and

$$\overline{\Delta I_{gr}^2} = \overline{N_{gr}}\, \Delta i_d^2. \qquad (A.4)$$

As $\overline{N_{gr}} \gg 1$, the probability density $p(I)$ adheres to a normal law. The approach is given by

$$p(I) \approx \frac{1}{\sqrt{\overline{\Delta I_{gr}^2}}} \varphi^{(0)}(x) - \frac{\lambda_3}{3!\, \overline{\Delta I_{gr}^2}^{\,2}} \varphi^{(3)}(x) + \cdots, \qquad (A.5)$$

where

$$x = \frac{I - \overline{I_{gr}}}{\sqrt{\overline{\Delta I_{gr}^2}}} \quad \text{and} \quad \varphi^{(n)}(x) = \frac{1}{\sqrt{2\pi}} \frac{d^n}{dx^n} \exp\left\{-\frac{x^2}{2}\right\}. \qquad (A.6)$$

Higher-order terms tend towards zero as $\overline{N_{gr}} \gg 1$. In this case, the probability density approaches a normal distribution

$$p(I) \approx 1/\sqrt{2\pi \overline{\Delta I_{gr}^2}} \cdot \exp\left\{-\frac{(I - \overline{I_{gr}})^2}{2\, \overline{\Delta I_{gr}^2}}\right\}. \qquad (A.7)$$

## Appendix B. Overlap of succeeding g-r pulses for g-r shot noise in the bulk

The g-r pulses in Eq. (A.1) occur at a generation rate of $g_0 = N_T/\tau_T$. Correspondingly, the time between succeeding pulses is $\tau_T/N_T$. Replacing $\tau_T$ in Eq. (60) by $\tau_T/N_T$ and $\tau_s$ by $\tau_{gr}$, the probability for an overlap is obtained by

$$P_{ov} = \frac{N_T \frac{\tau_{gr}}{\tau_T}}{1 + N_T \frac{\tau_{gr}}{\tau_T}} \qquad (B.1)$$

applying for an arbitrary number of traps $N_T \geq 2$. Using Eq. (31), this equation can be expressed by

$$P_{ov} = \frac{\overline{N_{gr}}}{1 + \overline{N_{gr}}}. \qquad (B.2)$$

For $\overline{N_{gr}} \gg 1$, the probability for an overlap $P_{ov} \to 1$, i.e., a large number of conduction electrons contributing to the g-r process leads to a strong overlap of succeeding g-r pulses being accompanied by a normal distribution of $I_{gr}(t)$ (see Appendix A).

## Appendix C. Fortran program for the simulation of g-r shot noise due to a single trap

```
!       Fortran program for the simulation of g-r shot noise
        REAL random@
        REAL tT, tgr, th , ti, ta, te, PP(2,100), R(100)
        INTEGER A(100)
!       The program simulates 50 (= nmax/2) g-r pulses
        tT = 1.                          ! tT  = τ_T
        tgr = 0.25                       ! tgr = τ_gr
        ti = 0.
        ta = 0.
        n = 1
        nmax = 100
        OPEN(10,file='g_r_shot_noise.prn',access='sequential')
        k = 1
!       start of simulation
10      ti = -tT*log(random@())          ! generate random intervals tau_T
```

```
                ta = ta + ti                        ! ta = time of occurrence of a g-r pulse
                PP(1,k) = ta                        ! store ta
                PP(2,k) = +1                        ! increase switching amplitude by +1
                k = k + 1
                th = -tgr*log(random@())            ! generate random g-r relaxation time tau_gr
                te = ta + th                        ! te = time point of the end of the g-r pulse
                PP(1,k) = te                        ! store te
                PP(2,k) = -1                        ! decrease switching amplitude by -1
                if (k. eq. nmax) goto 20
                k = k + 1
                goto 10
20              continue
!               end of simulation
                do 30 n = 1,nmax
                R(n) = PP(1,n)                      ! shuffle PP(1,n) into R(n)
30              A(n) = 0                            ! clear A(n)
                call rsort@(A,R,nmax)               ! sorts the array R by setting pointers from 1 to N in the array A
                ia = 0
                write(10,*) 0.,ia                   ! write starting time 0, and set starting amplitude to 0
                do 40 n = 1,nmax
                m = A(n)                            ! write sorted time points of occurrence t
                t = PP(1,m)                         ! and corresponding switching amplitude ia to file
                write(10,*) t, ia
                ia = ia + PP(2,m)
40              write(10,*) t, ia
                CLOSE(10)
                END
```

## Declaration of conflicting interests

The author declares that he has no known conflicting financial interests or personal relationships that could have influenced the work reported in this paper.

## Acknowledgement

The author would like to thank Dr. Barbara Herzberger for proofreading the manuscript.

## References

[1] K.S. Ralls, W.J. Skocpol, L.D. Jackel, R.E. Howard, L.A. Fetter, R.W. Epworth and D.M. Tennant, *Discrete Resistance Switching in Submicrometer Silicon Inversion Layers: Individual Interface Traps and Low-Frequency (1/f ?) Noise. Phys. Rev. Lett.*, **52** (1984) 228-231.

[2] M.J. Uren, D.J. Day and M.J. Kirton, *1/f and random telegraph noise in silicon metal-oxide-semiconductor field-effect transistors*, Appl. Phys. Lett. 47,11 (1985) 1195-1197.

[3] M.J. Kirton, M.J. Uren, S. Collins, M. Schulz, A. Karmann and K. Scheffer, *Individual defects at the Si:SiO$_2$ interface. Semicond. Sci. Technol.*, 4 (1989) 1116.

[4] M.J. Kirton and M.J. Uren, *Noise in solid-state microstructures: New perspectives on individual defects, interface states and low-frequency ( 1 / f ) noise. Adv. in Phys.*, **38** (1989) 367-468.

[5] M.N. Mihaila, *Low-Frequency Noise in Nanomaterials and Nanostructures*. Chapter 18 in: Noise and Fluctuations Control in Electronic Devices. Edited by A. Balandin. American Scientific Publishers (2002).

[6] M. Mihaila, *1/f noise in nanomaterials and nanostructures: old questions in a new fashion*. In "Advanced experimental methods for noise research in nanoscale electronic devices", J. Sikula & M. Levinshtein (eds), Kluwer Academic,17-25 (2004).

[7] C. Leyris, F. Martinez, M. Valenza, A. Hoffmann, J. C. Vildeuil and F. Roy, "Impact of Random Telegraph Signal in CMOS Image Sensors for Low-Light Levels," *2006 Proceedings of the 32nd European Solid-State Circuits Conference*, Montreux, 2006, pp. 376-379, doi: 10.1109/ESSCIR.2006.307609.

[8] E. Simoen, B. Kaczer, M. Toledano-Luque and C. Claeys, *Random Telegraph Noise: From a Device Physicist's Dream to a Designer's Nightmare,* ECS Transactions, 39,1 (2011) 3-15.

[9] E. Simoen, *Random Telegraph Signals in Semiconductor Devices*. Institute of Physics (Bristol) (2016).

[10] M. Pelton, G. Smith, N.F. Scherer and R.A. Marcus, *Evidence for a diffusion-controlled mechanism for fluorescence blinking of colloidal quantum dots.* Proc. Natl. Acad. Sci. 104 (2007) 14249-14254.

[11] P. Frantsuzov, M. Kuno, B. Janko, R. A. Marcus, *Universal Emission Intermittency in Quantum Dots, Nanorods and Nanowires.* Nature Physics 4 (2008) 519.

[12] S. Machlup, Noise in Semiconductors: *Spectrum of a Two-Parameter Random Signal*. J. Appl. Phys. 25,3 (1954) 341-343.

[13] R. E. Burgess, *The Statistics of Charge Carrier Fluctuations in Semiconductors*. Proc. Phys. Soc. London B68 (1956) 1020-1027.
[14] K. van Vliet and J. Fassett, *Fluctuations due to Electronic Transport in Solids*. In "Fluctuation Phenomena in Solids" Ed. R. Burgess, p. 267. Academic, New York, 1965.
[15] V. Mitin, L. Reggiani and L. Varani, *Generation-Recombination Noise in Semiconductors*. Chapter 2 in *Noise and Fluctuations Control in Electronic Devices*. Edited by A. Balandin (2002) American Scientific Publishers.
[16] A. Papoulis. *Probability, Random Variables and Stochastic Processes*. International Student Edition. McGraw-Hill. Tokyo (1981).
[17] W.B. Davenport and W.L. Root, *An Introduction to the Theory of Random Signals and Noise*. McGraw-Hill, New York (1958).
[18] J.R. Carson, *The statistical energy-frequency spectrum of random disturbances*. Bell Syst. Techn. J. 10 (1931) 374-381.
[19] N. Campbell. *The study of discontinuous phenomena*. Proc. Camb. Phil. Soc. **15** (1909) 117-136.
[20] D.R. Cox and H.D. Miller, *The Theory of Stochastic Processes*. Chapman and Hall, London and New York (1965).
[21] E. Parzen, *Stochastic Processes*. Holden-Day (1962).
[22] Y.W. Lee, *Statistical Theory of Communication*. John Wiley & Sons, New York (1960).
[23] S.O. Rice, *Mathematical analysis of Random Noise*. American Telephone and Telegraph Co. New York (1944). Bell telephone system technical publications; Monograph B-1589.
[24] F. Grüneis, *An alternative form of Hooge's relation for 1/f noise in semiconductor materials*. Physics Letters A 383 (2019) 1401-1409.
[25] F. Grüneis, *Estimation of the lowest limit of 1/f noise in semiconductor materials*. Physics Letters A 384 (2020) 126145.

**Addendum**

to

# THE TRANSITION FROM GENERATION-RECOMBINATION NOISE IN BULK SEMICONDUCTORS TO DISCRETE SWITCHING IN SMALL-AREA SEMICONDUCTORS

FERDINAND GRÜNEIS
*Institute for Applied Stochastic*
*Rudolf von Scholtz Str. 4, 94036 Passau, Germany*
*Email: Ferdinand.Grueneis@t-online.de*

published in

Section 3.1 of the above mentioned publication is revised. Machlup's noise spectrum is investigated in more detail. We show that the statistics of the unobservable random telegraph signal (RTS) is manifested in the random succession of the g-r pulses.

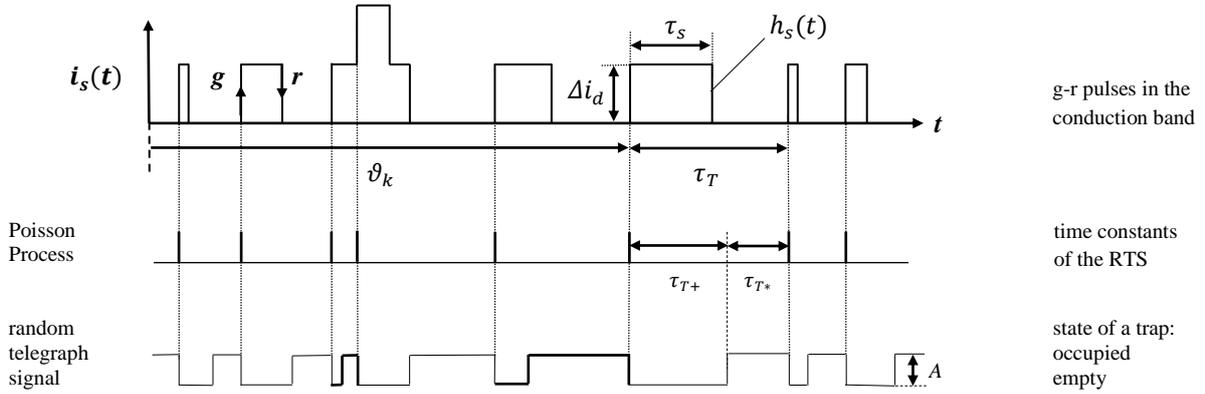

Fig. 5. Top: Current fluctuations $i_s(t)$ of a single trap represented by a random succession of g-r pulses $h_s(t)$ with switching amplitude $\Delta i_d$ and g-r relaxation time $\tau_s$. Middle: The underlying spike train is a Poisson process; each spike triggers an elementary g-r pulse $h_s(t)$. The time between successive generations is $\tau_T = \tau_{T+} + \tau_{T*}$. Bottom: Random telegraph signal representing an occupied or an empty state of a single trap. $A$ is the amplitude of the RTS. The bold RTS pulses indicate that an electron transferred from a neighboring trap has occupied the empty single trap.

### 6.1. *Machlup's noise spectrum for a single active trap*

For a single active trap Eq. (4) doesn't apply. Machlup [12] offers an alternative approach to the noise spectrum. In the following, we compare Machlup's noise spectrum with the g-r shot noise due to a single trap.

In line with Machlup's approach [12], we regard a single trap being either occupied by an electron or empty. The average lifetime of the occupied state is $\tau_{T*}$ and that of the empty state, $\tau_{T+}$ (see the middle of Fig. 5). The corresponding RTS is shown at the bottom of Fig. 5. Mathematically an (asymmetric) RTS is described by a two-state Markov process [20-22]. Machlup derived the power spectral density of a RTS by

$$S_{RTS}(f) = A^2 \frac{4}{(\tau_{T*}+\tau_{T+})\left[\left(\frac{1}{\tau_{T*}}+\frac{1}{\tau_{T+}}\right)^2 + (2\pi f)^2\right]}. \tag{37}$$

$A$ is the amplitude of the RTS (Fig. 5). For convenience we introduce the relaxation time of the RTS by

$$\frac{1}{\tau_{RTS}} \equiv \frac{1}{\tau_{T*}} + \frac{1}{\tau_{T+}}. \tag{38}$$

Also considering $\tau_T = \tau_{T*} + \tau_{T+}$ (Fig. 5), Eq. (37) is transformed into

$$S_{RTS}(f) = A^2 \frac{\tau_{RTS}^2}{\tau_T} \frac{4}{1+(2\pi f \tau_{RTS})^2}. \tag{39}$$

Machlup defines the empty and the occupied state by 0 and 1 respectively. Hence, the amplitude $A = 1$ reducing Eq. (39) to the spectral density (dimension 1/Hz) of the RTS.

In the following, we show that the statistics of the unobservable RTS is manifested in the random succession of the g-r pulses: for a single trap Eq. (4) reduces to

$$g_0 = \gamma N_{T*} \quad \text{and} \quad r_0 = \rho N_{T+}. \tag{40}$$

Like Eq. (37) it is symmetric under an exchange of the generation and recombination process. The master-equation approach leads to the single trap's g-r relaxation time by

$$\frac{1}{\tau_s} \equiv \gamma + \rho. \tag{41}$$

Obviously, $\tau_s \geq \tau_{gr}$ i.e., $\tau_s$ is larger than or equal to the g-r relaxation time in the bulk. Comparing Eq. (40) with Eq. (6) yields $\gamma = 1/\tau_{T*}$ and $\rho = 1/\tau_{T+}$ resulting in

$$\tau_{RTS} = \tau_s \tag{42}$$

i.e. the relaxation time of the unobservable RTS is identical with the empirically observable g-r relaxation time of a single trap[8]. Considering that $\tau_s$ is the lifetime of a g-r pulse with switching amplitude $A = \Delta i_d$, Eq. (39) is transformed into Machlup's noise spectrum (dimension $A^2$/Hz) for a single trap by

$$S_M(f) = \Delta i_d^2 \frac{\tau_s^2}{\tau_T} \frac{4}{1+(2\pi f \tau_s)^2}. \tag{43}$$

Applying Eqs. (12) and (41) leads to

$$\frac{\tau_s}{\tau_T} = f_T(1 - f_T) \tag{44}$$

which is shown in Fig. 6.a versus trap's occupancy. Substitution into Eq. (43) results in

$$S_M(f) = 4 f_T(1 - f_T) \Delta i_d^2 \frac{\tau_s}{1+(2\pi f \tau_s)^2}. \tag{45}$$

This form of Machlup's noise spectrum has been applied by several authors [1, 7].

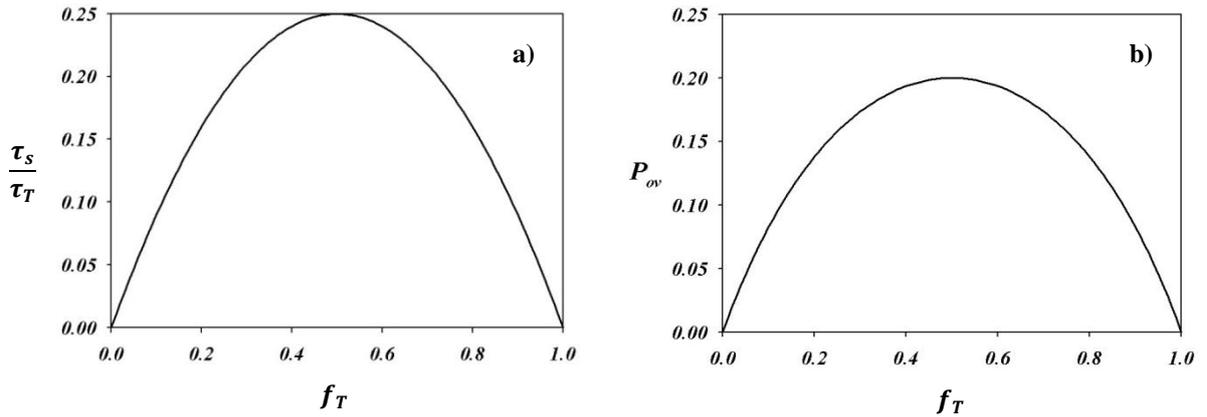

Fig. 6.a) The quotient $\tau_s/\tau_T$ and b) the probability of an

---

[8] Machlup equates the g-r relaxation time for a single trap with the lifetime of an empty trap, i.e. he defines $\tau_s \equiv \tau_{T+}$. Most researchers [1-9] agree with Machlup's interpretation. This interpretation, however, is inconsistent with above definitions of $\tau_s$ and $\tau_{RTS}$. For $\tau_{T*} \gg \tau_{T+}$ (i.e. for high occupancy $f_T \to 1$), Eq. (41) provides $\tau_s \approx \tau_{T+}$ approaching Machlup's definition for the g-r relaxation time of a single trap asymptotically.